\documentclass[preprint,prl,titlepage,preprintnumbers,amsmath,amssymb,letterpaper,superscriptaddress]{revtex4}

\usepackage{graphicx}
\usepackage{dcolumn}
\usepackage{bm}
\usepackage{times}
\usepackage{amsmath}
\usepackage{amsfonts}
\usepackage{amssymb}

\begin{document}

\title{Contamination of polymethylmethacrylate by organic quantum emitters}

\author{Andre Neumann$^{1,*}$, Jessica Lindlau$^{1,*}$, and Alexander H\"ogele}

\affiliation{Fakult\"at f\"ur Physik, Munich Quantum Center, and
Center for NanoScience (CeNS), Ludwig-Maximilians-Universit\"at
M\"unchen, Geschwister-Scholl-Platz 1, D-80539 M\"unchen, Germany}

\date{\today}

\begin{abstract}
We report the observation of ubiquitous contamination of
polymethylmethacrylate by organic molecules with optical activity
in the visible spectral range. Contamination sites of individual
solvent-specific fluorophores in thin films of
polymethylmethacrylate constitute fluorescence hot-spots with
quantum emission statistics and quantum yields approaching 30\% at
cryogenic temperatures. Our findings not only resolve prevalent
puzzles in the assignment of spectral features to various
nanoemitters in polymer matrices, they also identify means for
simple and cost-efficient realization of single-photon sources in
the visible spectral range.
\end{abstract}

%\keywords{Fluorescence spectroscopy, polymer matrix, organic
%fluorophores, single-photon emitters}

\maketitle

Embedding quantum emitters within chemically and electrostatically
inert polymer matrices such as polymethylmethacrylate (PMMA) is a
common approach to reduce the fluorescence (FL) intermittency
encountered by a broad class of photoactive nanoparticles
\cite{Nirmal1996,Frantsuzov2008} and molecular dyes
\cite{Basche1995} under ambient conditions, thus promoting stable
and enhanced FL \cite{Bradac2010,Ai2011}. However, contamination
of the polymer matrix by fluorescent constituents can result in
controversial assignment of spectral features. In some
spectroscopy experiments it has proven difficult to distinguish
between the FL stemming from quantum emitters, the polymer matrix,
or the supporting substrate \cite{Wang2015,Rabouw2016}. This is
not surprising given the challenge of unambiguous assignment of
the FL to its actual source for photoactive systems with low
quantum yields, or individual quantum emitters with high quantum
yields but inherently low absolute FL intensities.

In the visible spectral range, the realm of photoactive
nanoemitters includes single molecules \cite{Basche1997},
fluorescent nanodiamonds \cite{Aharonovich2011}, colloidal quantum
dots \cite{Klimov2007} and nanoplatelets \cite{Ithurria2011},
transition metal dichalcogenide quantum dots
\cite{Srivastava2015,He2015,Koperski2015,Chakraborty2015}, or
perovskite nanoplatelets \cite{Weidman2016}. The range of related
potential applications in light emitting, detecting, and harvesting
devices is as diverse as the specific details of the photophysics
of the underlying emitters. In absolute terms, however, and
depending on the radiative lifetime, some of these systems feature
low FL intensities despite high quantum yields, while others
suffer from reduced quantum yields due to optically inactive
lowest-lying dark states \cite{Basche1995,Efros1996,Zhang2015}
with strongly inhibited FL at cryogenic temperatures. Irrespective
of the actual reason for low intensity, any contamination of the
relevant FL by photoemissive substrates or matrices is clearly
detrimental to both fundamental studies of nanoemitters and their
related applications.

In the following, we present a comprehensive study targeting a
quantitative analysis of the FL in the visible spectral range
arising from a thin film of PMMA on various dielectric substrates.
Surprisingly, we find that PMMA films prepared by standard
solution-deposition procedures exhibit optical activity in the
visible both at room and cryogenic temperatures. However, the FL
is not a characteristic feature of the PMMA itself. It rather
stems from fluorescent contaminants in the PMMA matrix that we
ascribe to solvent residuals with specific FL intensity and
spectra. For individual fluorescent contaminants, pronounced and
spectrally stable zero-phonon lines (ZPLs) with red-shifted
vibronic satellites and highly non-classical emission statistics
emerge as a generic feature at cryogenic temperatures. At room
temperature, thermal broadening of both the ZPL and the
vibrational sidebands gives rise to a characteristic three-peak
spectrum that can be mistaken for phonon replica of silica color
centers \cite{Rabouw2016} or subject to other interpretations
\cite{Mason1998,English2002,Martin2008,Chizhik2009,Wang2009,Kusova2010,Schmidt2012}.

%%%%%%%%%%%%%%%%%%%%%%%%%%%%%%%%%%%%%%%%%% FIG 1 %%%%%%%%%%%%%%%%%%%%%%%%%%%%%%%%%%%%%%%%%%
\begin{figure}[t]
\begin{center}
\includegraphics[scale=1.1]{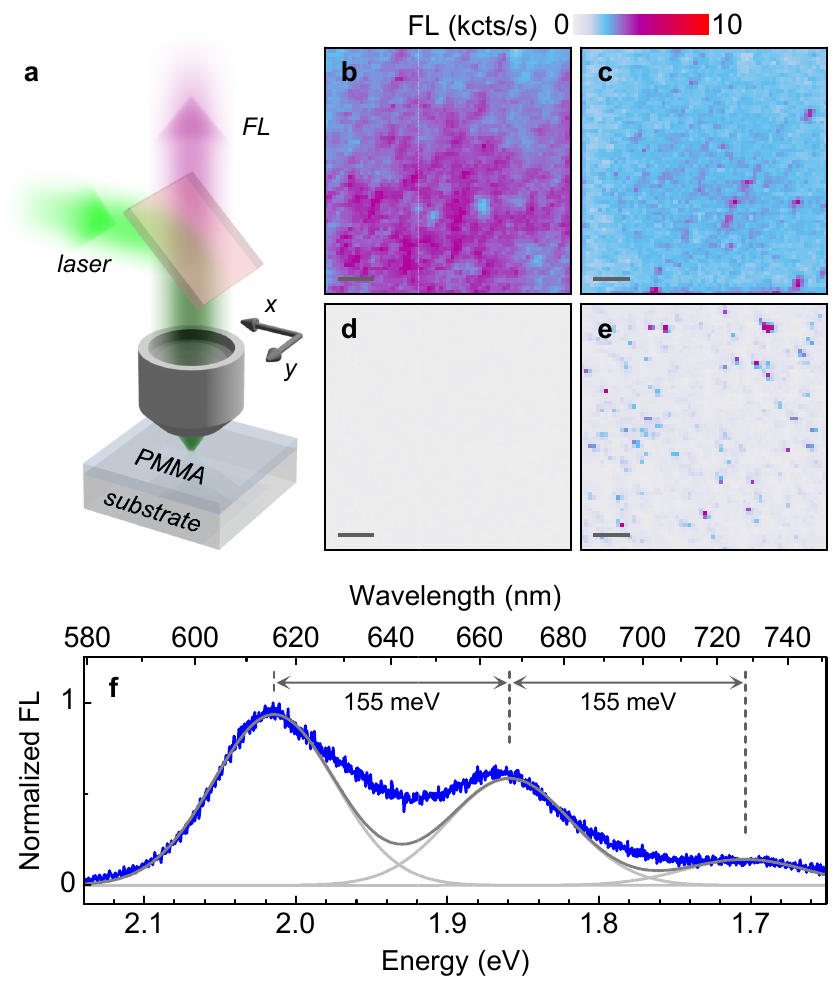}
\caption{\textbf{a}, Schematic of the experiment: fluorescence
from dielectric substrates coated with a thin film of PMMA was
studied with confocal raster-scan imaging and spectroscopy.
Raster-scan fluorescence intensity maps of the surface of fused
silica \textbf{b}, after sonication in acetone and isopropanol,
\textbf{c}, additional sonication in deionized water, and
\textbf{d}, oxygen plasma treatment. \textbf{e}, Fluorescence
image of plasma-cleaned fused silica with a thin film of PMMA
formulated in anisole. All maps were recorded with laser
excitation at $532$~nm and $130$~kW/cm$^2$; the scale bars are
$3~\mu$m. \textbf{f}, Fluorescence spectrum of a typical hotspot
in PMMA with a fit by three Gaussian peaks with full-width at
half-maximum linewidths of $90$~meV and equidistant energy
spacings of $155$~meV.} \label{fig1}
\end{center}
\end{figure}
%%%%%%%%%%%%%%%%%%%%%%%%%%%%%%%%%%%%%%%%%%%%%%%%%%%%%%%%%%%%%%%%%%%%%%%%%%%%%%%%%%%%%%%%%%%

The basics of our experiment are illustrated in Fig.~\ref{fig1}a.
We performed FL spectroscopy in a home-built optical microscope to
study sample-specific emission in the spectral range of $560-
770$~nm excited with a continuous-wave laser at $532$~nm, a
wavelength frequently used to excite FL in the visible. By
raster-scanning the sample with respect to fixed
diffraction-limited confocal excitation and collection spots, we
acquired maps of FL intensity as in Fig.~\ref{fig1}b~-~e with a
single photon counting avalanche photodiode (APD), and
hyperspectral maps with spectrally dispersed FL as in
Fig.~\ref{fig1}f recorded at each raster-scan pixel for spectral
analysis of individual emission hotspots. The studies were
complemented by time-correlated FL, second-order FL coherence and
FL excitation spectroscopy experiments performed either at room
temperature or at the cryogenic temperature of $3.1$~K.

In the first stage of the experiments we studied the FL
characteristics of bare dielectric substrates. It has been argued
recently that silica-based substrates host intrinsic fluorescent
centers with sizable FL intensity in the visible
\cite{Rabouw2016}. Therefore, we first investigated the FL from
the surface of bare fused silica substrates exposed to different
cleaning procedures (see Methods for details on cleaning
protocols). Under ambient conditions and $250~\mu$W irradiation in
a full-width at half-maximum (FWHM) spot of $0.5~\mu$m we acquired
raster-scan FL maps shown in Fig.~\ref{fig1}b~-~d. For fused
silica sonicated subsequently in acetone and isopropanol according
to a common cleaning procedure we observed FL from the entire
sample surface with inhomogeneous intensity and an average APD
count rate of $\sim 4$~kcts/s (Fig.~\ref{fig1}b). After an
additional sonication step in deionized water the level of FL
decreased to an average of $\sim 2$~kcts/s away from hotspot
emission with $\sim 4$~kcts/s (Fig.~\ref{fig1}c). Most remarkably,
additional treatment with oxygen plasma suppressed the FL from the
silica surface below the dark count rate of the APD
(Fig.~\ref{fig1}d). This set of data, consistently observed for
quartz and sapphire substrates subjected to oxygen plasma
treatment (see the Supplementary Information for
substrate-specific FL maps), clearly establishes the absence of
intrinsic FL defects on silica substrates. Moreover, it provides a
first hint at the source of the FL as stemming from organic
surface contaminants that do not withstand oxygen plasma
treatment.

For the second experimental stage we prepared substrates free of
FL background and covered them by spin-coating with PMMA dissolved
in anisole. On a silica substrate with $200$~nm of PMMA, we
observed the reappearance of fluorescent hotspots with intensities
of up to $\sim 6$~kcts/s on a background of $\sim 0.5$~kcts/s
(Fig.~\ref{fig1}e) under measurement conditions identical to those
of Fig.~\ref{fig1}b~-~d. Similar results were found for as-deposited
and thermally cross-linked PMMA films fabricated from anisole
solutions (see Methods for sample details). For most
hotspots, the FL was spatially localized to the
diffraction-limited spot and characterized by room temperature
spectra as in Fig.~\ref{fig1}f. The spectrum with maximum FL at
$2.02$~eV ($614$~nm) can be reproduced with some success by three
overlapping Gaussian peaks with FWHM linewidths of $90$~meV,
equidistant separations of $155$~meV, and intensities that reduce
with decreasing emission energy (grey solid lines in
Fig.~\ref{fig1}f). An explanation for the mismatch between this
simplistic model fit and the actual spectrum pending, we point out
its striking similarity to the spectra ascribed earlier to various
sources
\cite{Mason1998,English2002,Martin2008,Wang2009,Chizhik2009,Schmidt2012,Wang2015,Rabouw2016}.
Moreover, it exhibits a remarkable correspondence with the spectra
of individual dyes in PMMA \cite{Trautman1994, Macklin1996},
providing a second hint to hydrocarbon molecules as a source for
misinterpretation and establishing a link to the visionary
association made between the spectra of non-blinking colloidal
quantum dots \cite{Wang2009} and organic dyes \cite{Orrit2009}.

%%%%%%%%%%%%%%%%%%%%%%%%%%%%%%%%%%%%%%%%%% FIG 2 %%%%%%%%%%%%%%%%%%%%%%%%%%%%%%%%%%%%%%%%%%
\begin{figure}[t]
\begin{center}
\includegraphics[scale=1.1]{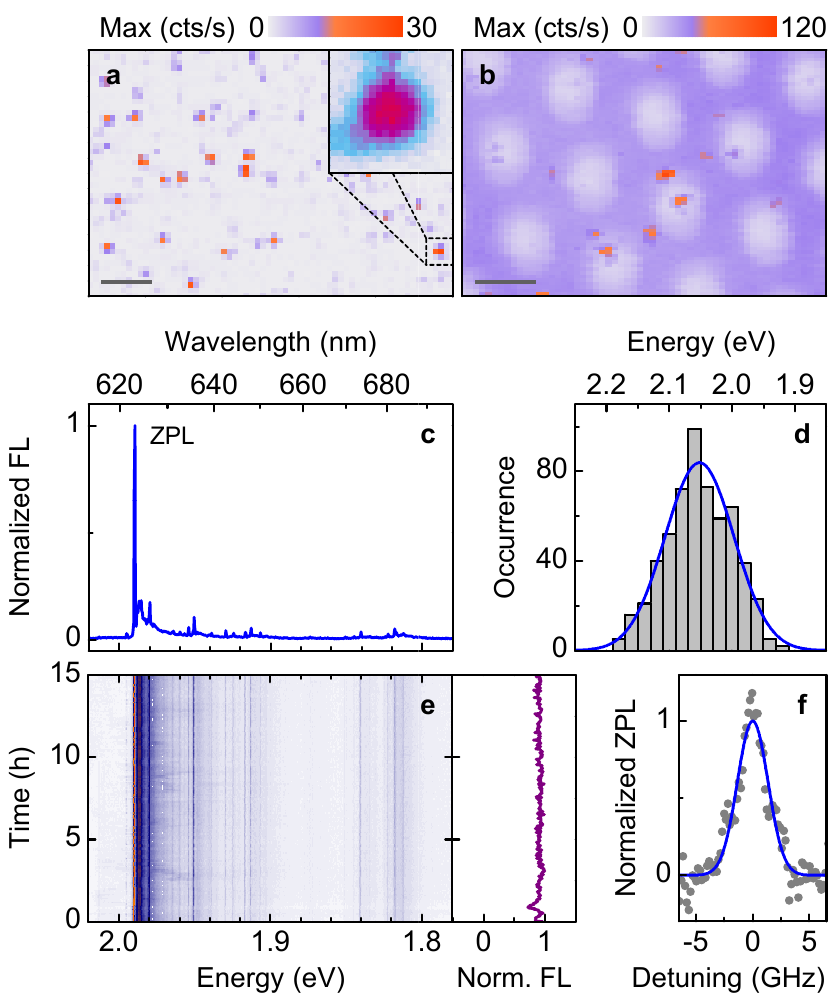}
\caption{ \textbf{a} and \textbf{b}, Cryogenic fluorescence peak
intensity maps of PMMA on a fused silica substrate and a
perforated silicon nitride membrane, respectively (recorded with
$532$~nm laser excitation at $26$ and $65$~kW/cm$^2$; the scale
bars are $3~\mu$m). The inset in \textbf{a} shows a $5\times$ zoom
to the hotspot delimited by the dashed box. The grey circular
areas in \textbf{b} are regions of freestanding PMMA. \textbf{c},
Normalized fluorescence spectrum of a typical hotspot with an
intense zero-phonon line (ZPL) and redshifted satellites.
\textbf{d}, Spectral distribution of the zero-phonon line of about
$600$ hotspots (the blue line is a Gaussian fit). \textbf{e},
Temporal evolution of the fluorescence spectrum (left) and
normalized intensity (right) over $15$~h for the hotspot with
spectrum in \textbf{c}. \textbf{f}, Zero-phonon line spectrum
resolved with a scanning Fabry-P\'{e}rot etalon (grey circles).
The Gaussian fit (blue line) yields an inhomogeneous linewidth of
$3.18 \pm 0.13$~GHz. All data were recorded at $3.1$~K.}
\label{fig2}
\end{center}
\end{figure}
%%%%%%%%%%%%%%%%%%%%%%%%%%%%%%%%%%%%%%%%%%%%%%%%%%%%%%%%%%%%%%%%%%%%%%%%%%%%%%%%%%%%%%%%%%%

To elucidate the correspondence between the FL hotspots found at
room temperature in thin films of PMMA and the spectral signatures
of organic molecules we carried out spectroscopy studies at the
cryogenic temperature of $3.1$~K. Fig.~\ref{fig2}a and b show
representative cryogenic FL maps of PMMA films on a fused silica
substrate and a perforated silicon nitride membrane, respectively.
Both maps were acquired in the hyperspectral mode by recording
spectrally dispersed FL with a nitrogen-cooled CCD and
color-coding its maximum intensity at each raster-scan pixel. Note
the conceptual difference to the raster-scan maps recorded with
APDs: hyperspectral mapping emphasizes emitters with sharp FL
peaks over spectrally broad FL background. Again, we found
spatially localized emission from diffraction-limited hotspots
(inset of Fig.~\ref{fig2}a) analogous to our room temperature
experiments. A few hotspots in Fig.~\ref{fig2}b (with up to
$120$~cts/s) clearly stem from PMMA regions suspended over holes
which can be unambiguously distinguished from the silicon nitride
membrane by the respective FL background (grey and blue areas of
the map correspond to intensities of $10$ and $50$~cts/s,
respectively). This observation confirms once more that the PMMA
film rather than the substrate is the actual host of FL hotspots.

A characteristic cryogenic FL spectrum of a hotspot in PMMA is
shown in Fig.~\ref{fig2}c. It features a narrow and intense peak,
which we label as ZPL, accompanied by weak red-shifted satellites.
More than $60\%$ of localized emission sites exhibited similar
spectral characteristics at low temperature. Within this group of
emitters with spectrometer-limited ZPLs, $94\%$ of hotspots
constitute the class of emitters with a ZPL centered around
$2.05$~eV emission energy ($605$~nm emission wavelength). The
corresponding normal distribution of the ZPL energy is shown in
Fig.~\ref{fig2}d, where the blue solid line is a Gaussian fit to
the histogram with a FWHM of $130$~meV. The remaining $6\%$ of the
single-site emitters with intense FL were characterized by two
sharp ZPLs (see Supplementary Information for the corresponding
normal distribution of emission energies) accompanied by
red-shifted sidebands.

All spectra were remarkably stable over time without significant
FL intermittence during the course of observation of $15$~h
(Fig.~\ref{fig2}e) and beyond. Throughout the temporal evolution,
the ZPL remained spectrometer-limited to one pixel of the CCD
corresponding to an upper bound on the FWHM linewidth of
$200~\mu$eV. A high-resolution spectrum recorded with a scanning
Fabry-P\'{e}rot etalon suggests that spectral wandering broadens the
ZPL on sub-minutes timescale to an inhomogeneous peak with a FWHM
of $3.18\pm0.13$~GHz (Fig.~\ref{fig2}f). These spectral signatures
find their correspondence in the studies of hydrocarbon
fluorophores embedded in a polymer host matrix
\cite{Kettner1994,Kozankiewicz1994,Walser2009}. Within this
framework, low-temperature FL of single molecules is characterized
by a spectrally narrow ZPL associated with the principal
electronic transition \cite{Diehl2010} and sidebands stemming from
Franck-Condon transitions between vibronically dressed molecular
electronic states \cite{Tchenio1993,Myers1994}. Stabilized in
PMMA, single molecules exhibit FL with low intermittency and
ZPLs limited by spectral diffusion to $\sim 1$~GHz
\cite{Kozankiewicz1994,Walser2009a}. The red-shifted satellites of the ZPL are
equally well pronounced in vibronic spectroscopy
\cite{Tchenio1993a} of molecules with characteristic vibrational
degrees of hydrocarbon complexes.

%%%%%%%%%%%%%%%%%%%%%%%%%%%%%%%%%%%%%%%%%% FIG 3 %%%%%%%%%%%%%%%%%%%%%%%%%%%%%%%%%%%%%%%%%%
\begin{figure}[t]
\begin{center}
\includegraphics[scale=1.1]{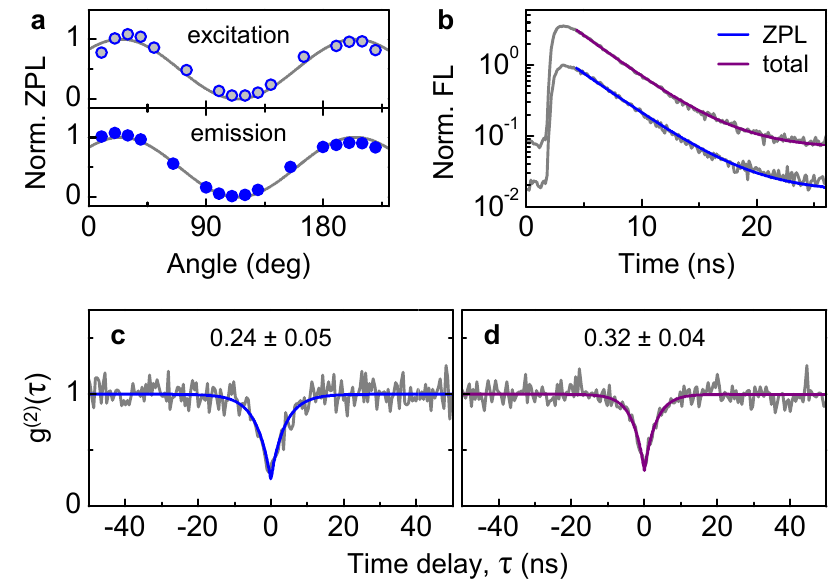}
\caption{\textbf{a}, Polarization characteristics of the
zero-phonon line: normalized intensity as a function of the
rotation angle of a linear polarizer in excitation and detection
(top and bottom panels, respectively; the grey lines represent the
same functional form of the fits to the data). \textbf{b},
Time-correlated decay of the zero-phonon line and the total
fluorescence (grey traces) of a single hotspot, and corresponding
monoexponential fits with decay constants of $3.8$ and $3.6$~ns
(blue and purple traces). The plots were offset for clarity.
\textbf{c} and \textbf{d}, Second-order coherence function
$g^{(2)}(\tau)$ recoded with and without spectral selection of the
zero-phonon line, respectively. The blue and purple lines are fits
to the data with multi-photon probabilities of $0.24 \pm 0.05$ and
$0.32 \pm 0.04$. All data were recorded at $3.1$~K with excitation
at $532$~nm.} \label{fig3}
\end{center}
\end{figure}
%%%%%%%%%%%%%%%%%%%%%%%%%%%%%%%%%%%%%%%%%%%%%%%%%%%%%%%%%%%%%%%%%%%%%%%%%%%%%%%%%%%%%%%%%%%

The set of data in Fig.~\ref{fig3} further substantiates the
correspondence. With polarization-resolved measurements shown in
Fig.~\ref{fig3}a we confirmed the dipolar character associated
with the molecular transition of the ZPL \cite{Macklin1996}. The
orientation of the absorption and emission axes measured with
linearly polarized excitation and detection, respectively, were
determined as collinear within our experimental precision.
Furthermore, time-correlated measurements of Fig.~\ref{fig3}b
revealed the characteristic FL decay dynamics of molecules on
nanoseconds timescale \cite{Green2015}. The single-exponential
lifetimes of $3.8$ and $3.6$~ns for the ZPL within a spectral
window of $60$~meV and the total FL intensity, respectively, were
the same within the temporal resolution of $\sim 0.3$~ns in our experiments,
identifying red-shifted sidebands as vibronic ZPL replicas.
Finally, single photon emission statistics as a hallmark of
single-molecule FL \cite{Basche1992, Lounis2000} are presented in
Fig.~\ref{fig3}c and d. With photon correlation spectroscopy we
observed pronounced photon antibunching in the normalized
second-order coherence function $g^{(2)}(\tau)$ at zero time delay
for both the FL within a band-pass interval of $60$~meV around the
ZPL (with $g^{(2)}(0) =0.24 \pm 0.05$ in Fig.~\ref{fig3}c) and the
full FL spectrum without spectral filtering ($g^{(2)}(0)=0.32 \pm
0.04$ in Fig.~\ref{fig3}d). Thus, within the uncertainty of our
measurement, we can rule out simultaneous photon emission into the
ZPL and the sideband spectrum associated with the vibronic ZPL
satellites.

%%%%%%%%%%%%%%%%%%%%%%%%%%%%%%%%%%%%%%%%%% FIG 4 %%%%%%%%%%%%%%%%%%%%%%%%%%%%%%%%%%%%%%%%%%
\begin{figure}[t]
\begin{center}
\includegraphics[scale=1.1]{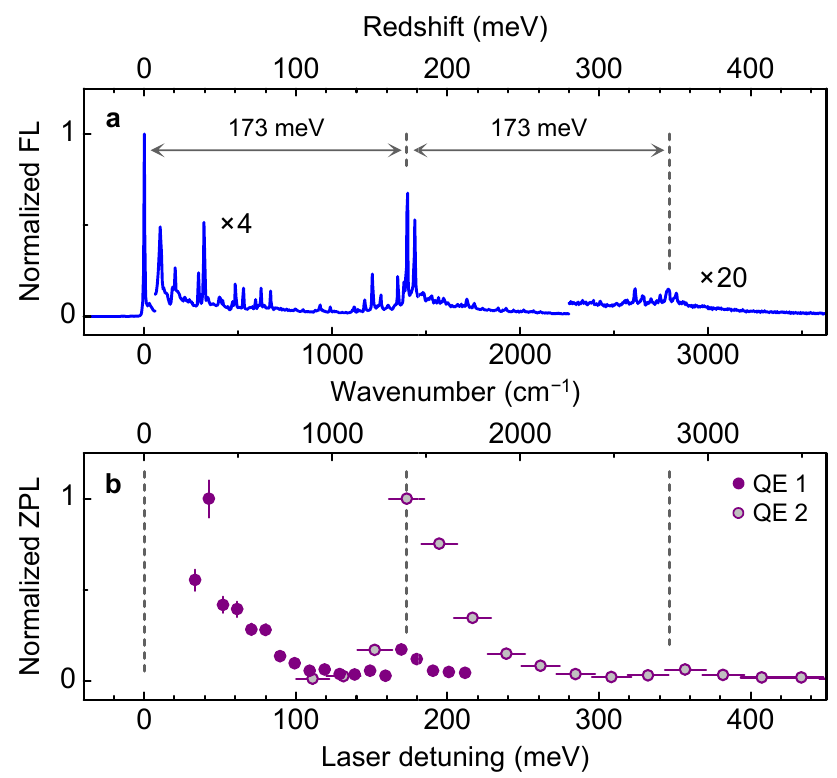}
\caption{\textbf{a}, Normalized fluorescence spectrum of a typical
hotspot recorded with excitation at $532$~nm and plotted as a
function of spectral redshift from the zero-phonon line at
$2.134$~eV. The intensity was magnified by a factor of $4$ ($20$)
above $7$~meV ($280$~meV) redshift to highlight two groups of
phonon sidebands with a spacing of $173$~meV. \textbf{b},
Normalized zero-phonon line intensity as a function of excitation
laser detuning for two quantum emitters (QE) with spectral
features as in \textbf{a} recorded within excitation wavelength
intervals of $538-583$ and $485-555$~nm (solid and open circles,
respectively). All data were recorded at $3.1$~K.} \label{fig4}
\end{center}
\end{figure}
%%%%%%%%%%%%%%%%%%%%%%%%%%%%%%%%%%%%%%%%%%%%%%%%%%%%%%%%%%%%%%%%%%%%%%%%%%%%%%%%%%%%%%%%%%%

Having identified the fluorescent hotspots in PMMA as single
fluorescent molecules, we utilized vibrationally resolved FL
spectroscopy \cite{Moerner1989, Tchenio1993a, Myers1994} to shed
light on their molecular nature. Fig.~\ref{fig4}a shows a spectrum
of a hotspot that is representative for fluorescent contaminants
in PMMA prepared with anisole as solvent. A series of
low-frequency vibrational modes contributes to the sidebands below
$80$~meV ($645~$cm$^{-1}$), followed by a group of replicas around
$173$~meV ($1395~$cm$^{-1}$) and a weaker satellite group around
$346$~meV ($2790~$cm$^{-1}$). The latter is in fact a second
harmonic of the preceding group as confirmed by correlation
analysis between all individual peaks of the two groups upon a
spectral shift by $173$~meV. All main vibrational features in
emission have their broadened counterpart resonances in
absorption, as demonstrated by the FL excitation spectra in
Fig.~\ref{fig4}b recorded for two typical emitters with different
ZPL energies as a function of laser energy detuning at constant
excitation power. For both quantum emitters of Fig.~\ref{fig4}b,
the absorption is enhanced whenever the laser detuning with
respect to the ZPL matches the energy of the vibronic sidebands
(the dashed lines in Fig.~\ref{fig4} emphasize the correspondence
between the resonances in emission and absorption).

The vibrationally resolved spectrum of Fig.~\ref{fig4}b is typical
for fluorescent molecules in PMMA films from anisole-based
solutions. It exhibits a striking similarity with the cryogenic FL
of anthracene characterized by a ZPL in the ultraviolet (around
$3$~eV) and a pronounced vibronic satellite group around
$1400$~cm$^{-1}$ redshifts \cite{Helfrich1965}. The according
vibrational degrees of freedom are related to the intramolecular
stretching of adjacent carbon bonds in polycyclic aromatic
hydrocarbons \cite{Myers1994}. The observation of the ZPL emission
in the visible (around $2$~eV) from anisole-based PMMA suggests
that the optical activity of solvent-related contaminants in such
films stems from acene chains such as pentacene, or from
anthracene-derived dyes such as alizarin.

\begin{figure}[t!]
\begin{center}
\includegraphics[scale=1.1]{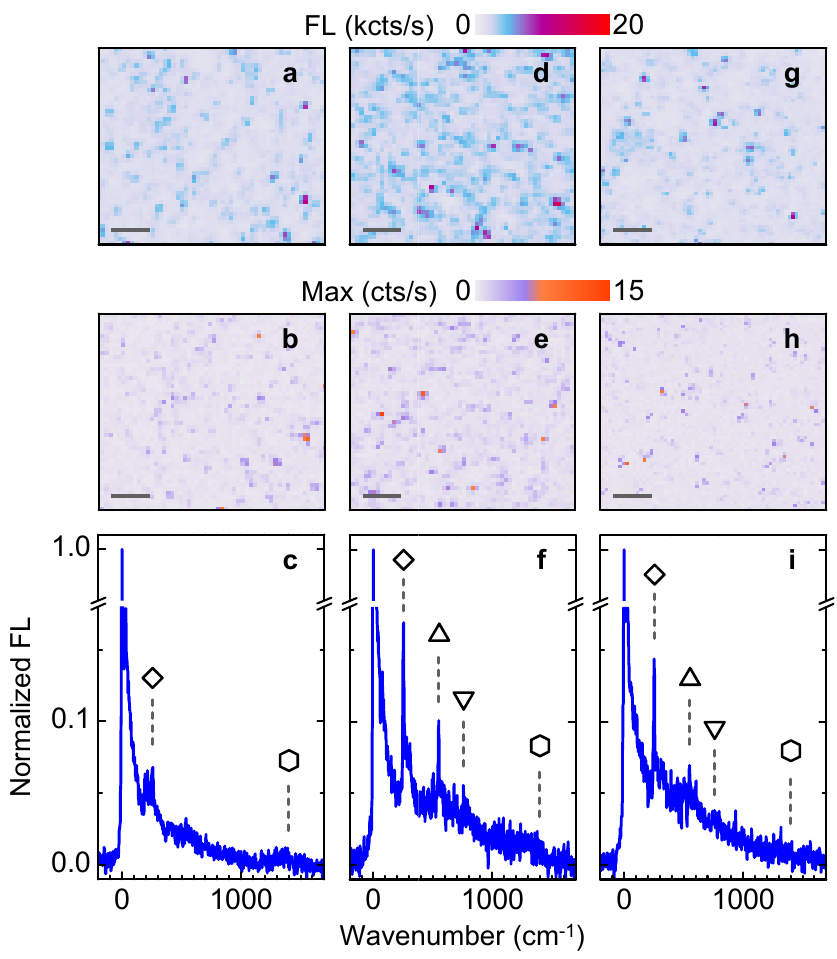}
\caption{\textbf{a}, \textbf{b}, Raster-scan maps of integrated
fluorescence intensity and spectrally dispersed fluorescence
maxima, respectively, for PMMA dissolved with chlorobenzene.
\textbf{d}, \textbf{e}, and \textbf{g}, \textbf{h}, Same for PMMA
films prepared by solution in methyl isobutyl ketone and toluene,
respectively. All scale bars are $3~\mu$m. \textbf{c}, \textbf{f},
and \textbf{i} Average fluorescence spectra of $25$ most intense
hotspots in PMMA dissolved in chlorobenzene, methyl isobutyl
ketone, and toluene, respectively, displayed as a function of the
redshift from the zero-phonon line. The symbols indicate specific
vibronic modes of hydrocarbon-based molecular dyes. All data were
recorded at $3.1$~K with laser excitation at $532$~nm and
$52~$kW/cm$^2$ on thermally annealed PMMA films of $200$~nm
thickness on oxygen plasma-cleaned fused silica.} \label{fig5}
\end{center}
\end{figure}

We applied vibrational FL spectroscopy to hotspots in PMMA films
derived from other solvents (see Methods for sample preparation
details). As highlighted by the raster-scan maps of
Fig.~\ref{fig5}, the areal density and the FL intensity of
hotspots in PMMA films formed with chlorobenzene (Fig.~\ref{fig5}a,
b), methyl isobutyl ketone (Fig.~\ref{fig5}c, d) and toluene
(Fig.~\ref{fig5}g, h) were similar to anisole-based PMMA
characteristics (Fig.~\ref{fig1}e and Fig.~\ref{fig2}a). The
vibronic signatures, however, showed significant differences.
Fig.~\ref{fig5}c, f, and i show normalized average spectra of $25$
brightest fluorophore contaminants in PMMA films prepared with
different solvents (see the Supplementary Information for the
corresponding average spectrum of anisole-based PMMA).

%%%%%%%%%%%%%%%%%%%%%%%%%%%%%%%%%%%%%%%%%% FIG 5 %%%%%%%%%%%%%%%%%%%%%%%%%%%%%%%%%%%%%%%%%%
%%%%%%%%%%%%%%%%%%%%%%%%%%%%%%%%%%%%%%%%%%%%%%%%%%%%%%%%%%%%%%%%%%%%%%%%%%%%%%%%%%%%%%%%%%%

The spectrum of a typical hotspot in chlorobenzene-based PMMA
features a low-frequency vibronic band around $250$~cm$^{-1}$ and
a pronounced high-frequency band around $1400$~cm$^{-1}$ discussed
earlier (as indicated by the diamond and the hexagon
Fig.~\ref{fig5}c). In contrast, the vibronic FL characteristics of
hotspots in PMMA films formed with methyl isobutyl ketone and
toluene solutions (Fig.~\ref{fig5}f and i, respectively) exhibit
additional vibrational signatures at $548$ and $757$~cm$^{-1}$.
The vibronic modes, labelled with diamonds and upper and lower
triangles in Fig.~\ref{fig5}c, f, and i are characteristic of
rylene dyes composed of naphthalenes. While the low-frequency mode
(diamonds in Fig.~\ref{fig5}c, f, and i) is close to that of the
long axis stretch of a terrylene molecule, the higher-frequency
modes (upper and lower triangles in Fig.~\ref{fig5}f and i) are
consistent with the short axis stretch and ring deformation of
outer naphthalenes, respectively \cite{Myers1994}. Note that
naphthalene-related bands of rylene dyes are only very weakly
expressed in the averaged vibronic FL spectra observed in anisole-
and chlorobenzene-based PMMA films (Fig.~\ref{fig4}a and
Fig.~\ref{fig5}c).

In addition to solvent-specific differences in the
spectra of fluorescent hotspots in PMMA, vibrationally resolved FL
spectroscopy identifies the normal modes of aromatic hydrocarbons
around $170$~meV ($1400$~cm$^{-1}$) as a generic feature of FL
contaminants at low temperatures. At elevated temperatures, these
modes develop into broad vibronic satellites (see the
Supplementary Information for FL spectra at different
temperatures) that accompany the FL from the thermally broadened
principal molecular transition. With this in mind, the
interpretation of the three-peak structure of the room-temperature
FL spectrum in Fig.~\ref{fig1}f as arising from an organic
fluorophore is straight forward. For an adequate modelling,
however, the contributions of all other vibrational modes must be
taken into account. The main corrections to the
inhomogeneous spectral profile of the ZPL and the vibronic modes
of polycyclic hydrocarbons will naturally appear on the low-energy
side of the peaks, where the fit with three Gaussians most
significantly deviates from the actual spectrum.

In concise terms, our comprehensive study of fluorescent spots,
ubiquitously present in PMMA films and on contaminated dielectric
substrates, leads to the conclusion that organic fluorophores are
the actual source of misinterpreted FL signatures. We estimate the
quantum yield of such organic quantum emitters to range from $\sim
5\%$ at room temperature up to $30\%$ at $3.1$~K (see the
Supplementary Information for the estimate of the quantum yield).
These values are not remarkably high, however, the corresponding
FL intensity can be significant in studies of photoactive systems
with reduced quantum yields in cryogenic or ambient environments.
In fact, we found the FL intensity of PMMA hotspots to be roughly
a third of the emission intensity of individual terrylenediimide
(TDI) molecules at cryogenic temperatures, and in many instances
even more intense than commercial radiant dyes at ambient
conditions. Given the present technological limitations to solvent
purity, it seems unlikely that contamination of PMMA and other
polymer matrices can be completely avoided in future experiments. On the
other hand, the abundance of stable quantum emitters in polymer
films could facilitate a range of fundamental studies and
technological developments relying on simple and cheap sources of
non-classical light.

%%%%%%%%%%%%%%%%%%%%%%%%%%%%%%%%%

{\it Methods:} All samples were prepared in a clean-room
environment. Unless stated otherwise, substrates were cleaned by
initial sonication in acetone (Technic, acetone Micropur VLSI) for
$5$~min, followed by isopropanol (Technic, propan-2-ol Micropur
VLSI) for $5$~min, and finally exposed for $1$~min to oxygen
plasma. Polymer covered samples were prepared by spin-coating
$\sim 10~\mu$l of PMMA onto oxygen plasma-treated fused silica
(CrysTec) and other dielectric substrates (quartz and sapphire).
An ellipsometer was used to adjust the spin-coating parameters for
a film thickness of $200$~nm. The films were obtained from
commercial PMMA formulated in anisole with a molecular weight of
$950$K (MicroChem, $950$PMMA A$4$ resist for electron-beam
lithography). The spin-coated PMMA film was left to dry at ambient
conditions. Optionally, the samples were baked at $180~^\circ$C
for $90$~s on a hot plate. The perforated silicon nitride membrane
(PELCO) of Fig.~\ref{fig2}b was drop-casted and baked to ensure
mechanical stability of freely suspended PMMA. Control experiments
were carried out with $4\%$ of $450$K PMMA resin (DuPont, Elvacite
2041) diluted in $96\%$ of chlorobenzene (Merck, 801791), methyl
isobutyl ketone (Technic, MIBK Micropur VLSI), or toluene
(Sigma-Aldrich, 179418).

FL imaging and spectroscopy measurements were performed with a
home-built confocal microscope coupled to single-mode fibers. Room-temperature experiments were
conducted with an apochromatic objective with numerical aperture
(NA) of $0.82$ (attocube systems, LT-APO/VISIR/0.82) and an oil
immersion objective (Olympus, UPLFLN 100XOI2) with NA of $1.30$
for the data in Fig.~\ref{fig1}f. Cryogenic experiments were
carried out in a helium bath cryostat or a low-vibration
closed-cycle cryostat (attocube systems, attoDRY1000) with base
temperatures of $4.2$~K and $3.1$~K, respectively, using a
low-temperature apochromatic objective with NA of $0.65$ (attocube
systems, LT-APO/VIS/0.65).

Continuous wave excitation with a solid-state laser at $532$~nm
(CNI, MLL-III-532-50-1) was used in all experiments except for the
measurements of data in Fig.~\ref{fig3}b and Fig.~\ref{fig4}b.
All FL maps were recorded with circularly polarized excitation except for Fig.~\ref{fig2}b and Fig.~S1, where linearly polarized excitation was used.
Time-resolved FL data in Fig.~\ref{fig3}b were measured with
ps-excitation at $532$~nm. The FL excitation experiments of
Fig.~\ref{fig4}b were performed with an optical parametric
oscillator (Coherent, Mira-OPO with a FWHM spectral bandwidth of
$0.5$~nm) or a spectrally filtered supercontinuum laser (NKT
Photonics, SuperK EXW-$12$ with a FWHM spectral bandwidth of
$5.5$~nm). Single photon counting avalanche photodiodes
(PicoQuant, $\tau$-SPADs with dark count rates of $35$~cts/s and a
temporal resolution of $320$~ps) or a monochromator equipped with
a liquid nitrogen cooled CCD (PI, Acton SP-2558 and
Spec-10:100BR/LN with a spectral resolution of $200~\mu$eV and a gain setting of $4$~e$^-$/count) were
used for detection. The hyperspectral raster-scan maps in
Fig.~\ref{fig2}a and b were recorded in the spectral range of
$1.68 - 2.20$~eV. The data in Fig.~\ref{fig2}f were measured with
a home-built monolithic scanning Fabry-P\'{e}rot etalon with a
spectral resolution of $150$~MHz and a scanning rate of
$5.5$~MHz/s.

{\it Conflict of interest:} The authors declare no competing
financial interest.

{\it Acknowledgment.} We thank T.~Basch\'{e}, C.~Br\"{a}uchle,
I.~Gerhard, S.~G\"otzinger, K.~Karrai, J.~P.~Kotthaus,
E.~Lifshitz, J.~Lupton, G.~I.~Maikov, M. Pilo-Pais, I.~Pugliesi,
K.~Puschkarsky, E.~Riedle, J.~Tilchin and S.~E.~Beavan for helpful
discussions and useful input at various stages of the project,
P.~Maletinsky and S.~Thoms for providing samples with NV centers
in diamond and TDI molecules in PMMA, respectively, and
P.~Altpeter and R.~Rath for assistance in the clean-room. This
work was funded by the Volkswagen Foundation, the German-Israeli
Foundation for Scientific Research and Development (GIF), and the
Deutsche Forschungsgemeinschaft (DFG) Cluster of Excellence
Nanosystems Initiative Munich (NIM) with financial support from
LMUinnovativ and the Center for NanoScience (CeNS). A.~H.
acknowledges funding by the European Research Council under the
ERC grant agreement no. $336749$.

{\it Author Contributions:} $^{\star}$these authors contributed
equally to this work.

{\it Corresponding author:} alexander.hoegele@lmu.de

{\it Supplementary Information} provides details on sample
fabrication and experiments.

%
%\bibliographystyle{apsrev}
%\bibliography{scibib_bibliography}

%%%%%%%%%%%%%%%%%%%%%%%%%%%%%%%%%%%%%%%%%%%%%%%%%%
\cleardoublepage
\begin{center}
\section{\Large{SUPPLEMENTARY INFORMATION}}
\end{center}

\setcounter{page}{1} \setcounter{figure}{0}
\setcounter{equation}{0}

\subsection{Sample fabrication}\label{sample}
\vspace{-8pt} All samples were prepared in a clean-room
environment. Unless stated otherwise, substrates were cleaned by
initial sonication in acetone (Technic, acetone Micropur VLSI) for
$5$~min, followed by isopropanol (Technic, propan-2-ol Micropur
VLSI) for $5$~min, and finally exposed for $1$~min to oxygen
plasma. Polymer covered samples were prepared by spin-coating
$\sim 10~\mu$l of PMMA onto oxygen plasma-treated fused silica
(CrysTec) and other dielectric substrates (quartz and sapphire).
An ellipsometer was used to adjust the spin-coating parameters for
a film thickness of $200$~nm. The films were obtained from
commercial anisole based PMMA with a molecular weight of $950$K
(MicroChem, $950$PMMA A$4$ resist for electron-beam lithography).
The spin-coated PMMA film was left to dry at ambient conditions.
Optionally, the samples were baked at $180~^\circ$C for $90$~s on
a hot plate. The perforated silicon nitride membrane (PELCO) of
Fig.~1b from the main text was drop-casted and baked to ensure
mechanical stability of freely suspended PMMA. Control experiments
were carried out with $4\%$ of $450$K PMMA resin (DuPont, Elvacite
2041) diluted in $96\%$ of chlorobenzene (Merck, 801791), methyl
isobutyl ketone (Technic, MIBK Micropur VLSI), or toluene
(Sigma-Aldrich, 179418).

\subsection{Experimental setup}\label{setup}
\vspace{-8pt} A home-built confocal microscope coupled to
single-mode fibers was used for FL imaging and spectroscopy.
Room-temperature experiments were performed with an apochromatic
objective with numerical aperture (NA) of $0.82$ (attocube
systems, LT-APO/VISIR/0.82) and an oil immersion objective
(Olympus, UPLFLN 100XOI2) with NA of $1.30$. Cryogenic experiments
were carried out in a helium dewar or a low-vibration closed-cycle
cryostat (attocube systems, attoDRY1000) with base temperatures of
$4.2$~K and $3.1$~K, respectively, using a low-temperature
apochromatic objective with NA of $0.65$ (attocube systems,
LT-APO/VIS/0.65). The FL was excited with a continuous wave
solid-state laser at $532$~nm (CNI, MLL-III-532-50-1), an optical
parametric oscillator (Coherent, Mira-OPO with a FWHM spectral
bandwidth of $0.5$~nm), or a spectrally filtered supercontinuum
laser (NKT Photonics, SuperK EXW-12 with a FWHM spectral bandwidth
of $5.5$~nm). Single photon counting avalanche photodiodes
(PicoQuant, $\tau$-SPADs with dark count rates of $35$~cts/s and a
temporal resolution of $320$~ps) or a monochromator equipped with
a liquid nitrogen cooled CCD (PI, Acton SP-2558 and
Spec-10:100BR/LN with a spectral resolution of $200~\mu$eV and a
gain setting of $4$~e$^-$/count) were used for detection. A
home-built monolithic scanning Fabry-P\'{e}rot etalon with a
spectral resolution of $150$~MHz was used for high-resolution
spectroscopy.

\subsection{Optical characterization of substrates}\label{substrates}
\begin{figure*}[hb!]
\begin{center}
\includegraphics[scale=1.1]{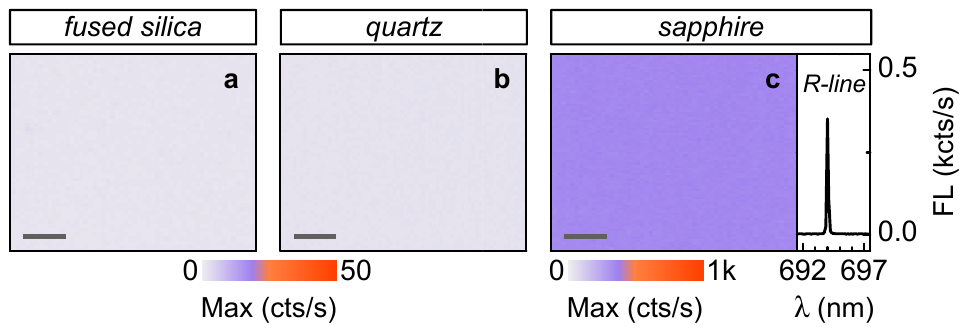}
\vspace{-15pt} \caption{\textbf{a}, \textbf{b}, and \textbf{c},
Raster-scan maps of spectrally dispersed fluorescence maxima for
oxygen plasma-treated fused silica, quartz, and sapphire
substrate, respectively. The hyperspectral maps were recorded
between $1.68$~eV and $2.20$~eV with a binning of $0.4$~meV at an
excitation power density of $130$~kW/cm$^2$. The temperature was
$3.1$~K. The fluorescence intensity levels in \textbf{a} and
\textbf{b} correspond to the readout noise of the CCD, the
homogeneous level of $1$~kcts/s in \textbf{c} corresponds to the
R-line emission at $\sim 694$~nm ($1.787$~eV) of Cr$^{3+}$ ions in
sapphire with a characteristic spectrum shown in the right panel.
All data were measured with a linearly polarized $532$~nm laser;
scale bars are $3~\mu$m.} \label{si_substrates}
\end{center}
\end{figure*}

\vspace{-8pt} Three dielectric substrates were studied in
hyperspectral raster-scan FL:  fused silica (Crystec), quartz
(Crystec, z-cut, $0001$ orientation) and sapphire (MaTecK, z-cut,
$0001$ orientation). Prior to cryogenic measurements the
substrates were exposed to oxygen plasma. Cryogenic raster-maps of
FL maxima are shown in Fig.~\ref{si_substrates}a, b, and c for
fused silica, quartz, and sapphire, respectively. The FL maximum
level was identical for fused silica and quartz
(Fig.~\ref{si_substrates}a and b) with count rates given by the
readout noise of the liquid nitrogen cooled CCD. The sapphire
substrate exhibited spatially homogeneous FL intensity stemming
from a sharp peak at $\sim 694$~nm ($1.787$~eV) of the R-line of
Cr$^{3+}$ ions in Al$_2$O$_3$.

%\vspace{8pt}
%%%%%%%%%%%%%%%%%%%%%%%% FIG %%%%%%%%%%%%%%%%%%%%%%%%%%%%%%%%%%%%%%%%%%%%%%%%%%%%%%%%%%%%%%
%%%%%%%%%%%%%%%%%%%%%%%%%%%%%%%%%%%%%%%%%%%%%%%%%%%%%%%%%%%%%%%%%%%%%%%%%%%%%%%%%%%%%%%%%%%

\subsection{Optical characterization of PMMA prepared with different solvents}\label{solvents}
\begin{figure*}[h!b!]
\begin{center}
\includegraphics[scale=1.1]{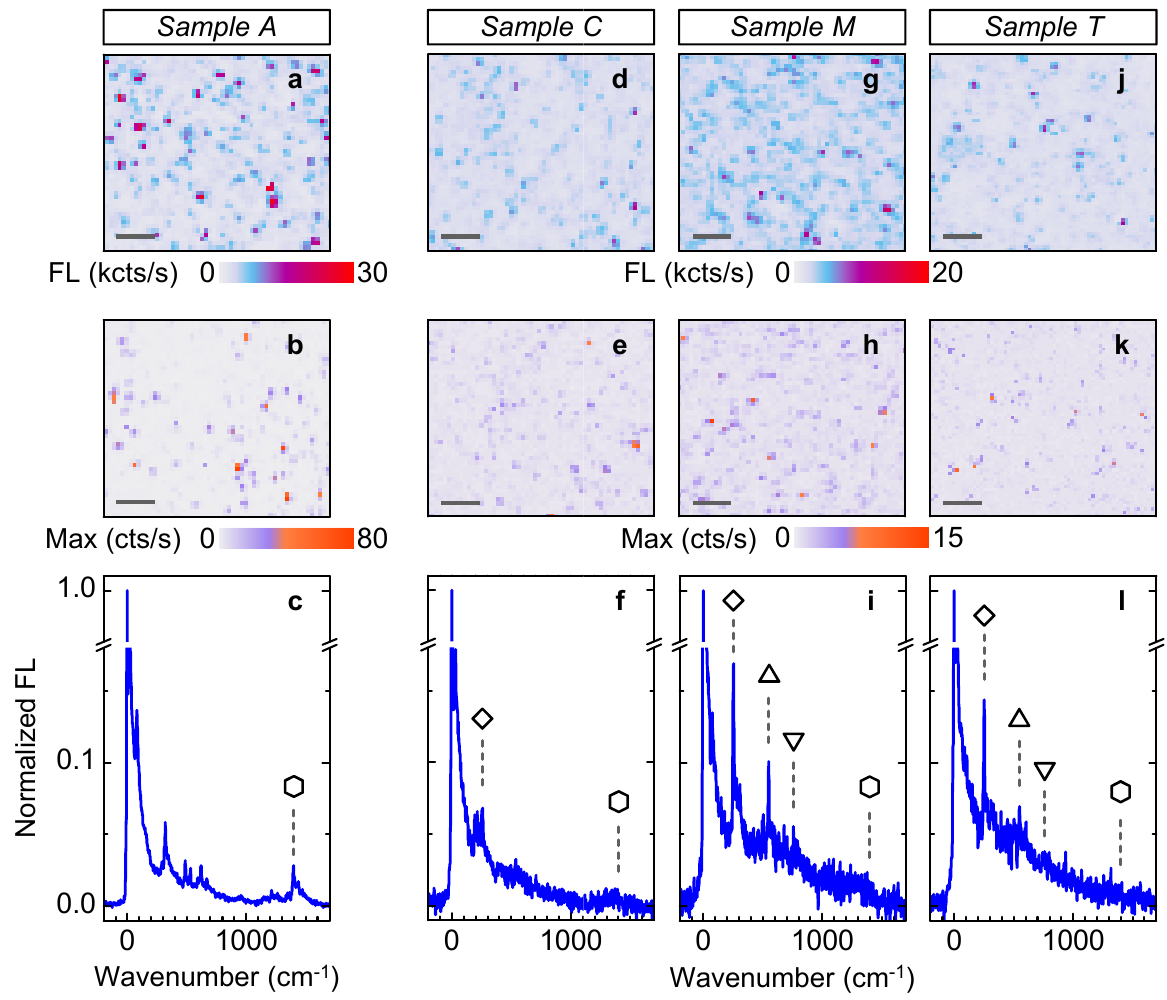}
\vspace{-10pt} \caption{\textbf{a}, and \textbf{b}, Raster-scan
maps of integrated fluorescence intensity and spectrally dispersed
fluorescence maxima, respectively, for PMMA dissolved with
anisole. \textbf{d} and \textbf{e}, \textbf{g} and \textbf{h},
\textbf{j} and \textbf{k}, Same for PMMA films prepared with
chlorobenzene, methyl isobutyl ketone, and toluene solutions,
respectively. \textbf{c}, Normalized ensemble spectrum of $25$
brightest fluorescence hotspots in anisole-based PMMA, plotted in
wavenumbers as a function of red-shift from the zero-phonon line.
\textbf{f}, \textbf{i}, and \textbf{l}, Same for PMMA formed with
chlorobenzene, methyl isobutyl ketone, and toluene, respectively.
Symbols indicate to hydrocarbon-specific vibronic modes of
molecular dyes (see main text). All data were measured on
thermally annealed PMMA films of $200$~nm thickness on oxygen
plasma-cleaned fused silica substrates at $3.1$~K. The samples
were excited at $532$~nm with a power density of $52~$kW/cm$^2$
and circular polarization. All scale bars are $3~\mu$m. The data
in \textbf{d}~-~\textbf{l} are the same as in Fig.~5 of the main
text.} \label{si_solvents}
\end{center}
\end{figure*}

\vspace{-8pt} Fig.~\ref{si_solvents} compares cryogenic FL
characteristics of PMMA films with $200$~nm thickness, prepared by
spin-coating mixtures of PMMA in different solvents onto oxygen
plasma-treated fused silica substrates. Four different solvents
were used: anisole (\mbox{sample A}), chlorobenzene (\mbox{sample
C}), methyl isobutyl ketone (\mbox{sample M}), and toluene
(\mbox{sample T}).

%\vspace{-20pt}
%%%%%%%%%%%%%%%%%%%%%%%% FIG %%%%%%%%%%%%%%%%%%%%%%%%%%%%%%%%%%%%%%%%%%%%%%%%%%%%%%%%%%%%%%

\subsection{Spectral characteristics of quantum emitters in anisole-based PMMA films}\label{temperature_power}
\begin{figure*}[ht!]
\begin{center}
\includegraphics[scale=1.1]{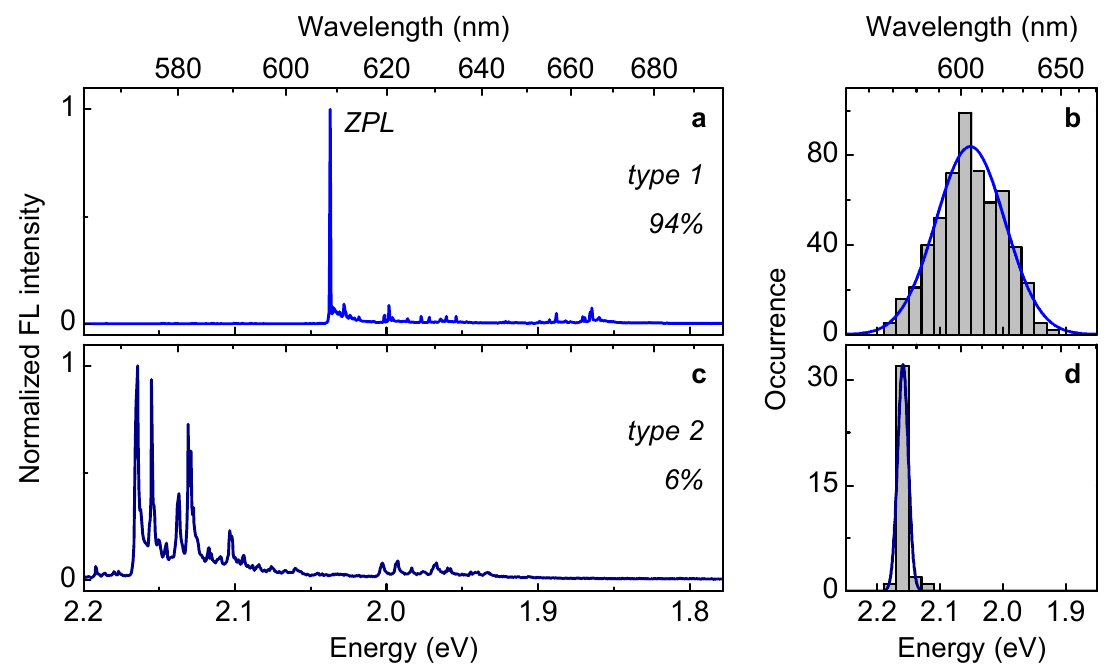}
\vspace{-15pt} \caption{\textbf{a}, Normalized fluorescence
spectrum of a hotspot with an intense zero-phonon line (ZPL) and
red-shifted satellites typical for type~1 quantum emitters in
anisole-based PMMA. \textbf{b}, Spectral distribution of the
zero-phonon line of about $600$ such hotspots (same as in Fig.~2d
in the main text). \textbf{c}, Representative normalized spectrum
of type~2 hotspot emitters featuring two intense and sharp
emission peaks with two intense vibronic satellites. \textbf{d},
Corresponding energy distribution of the blue-most peak in type~2
emission spectrum. The fraction of type~1 and emitters 2 is given
in percent in \textbf{a} and \textbf{c}, respectively. The blue
lines in \textbf{b} and \textbf{d} are normal distribution fits
with full-widths at half-maximum of $130$ and $20$~meV,
respectively. All data were measured at $3.1$~K and excitation at
$532$~nm.} \label{si_typetwo}
\end{center}
\end{figure*}

\vspace{-8pt} Representative spectra of most common hotspot
quantum emitters in PMMA films prepared with anisole solvent are
shown in Fig.~\ref{si_typetwo}. More than $60\%$ of localized
emission sites exhibited similar spectral characteristics at low
temperature. Within this group of emitters with
spectrometer-limited ZPLs, $94\%$ of hotspots constituted the
class of emitters with one sharp and intense zero-phonon line
(ZPL) and red-shifted vibronic replicas as in
Fig.~\ref{si_typetwo}a. The corresponding normal distribution of
the ZPL energy is shown in Fig.~\ref{si_typetwo}b. In contrast to
such emitters that we label here as type~1, $6\%$ of type~2
fluorescent hotspots exhibited two intense and sharp peaks as in
Fig.~\ref{si_typetwo}c with two intense satellites and a much
narrower spread in the energy of the blue-most peak (histogram in
Fig.~\ref{si_typetwo}d).

%%%%%%%%%%%%%%%%%%%%%%%% FIG %%%%%%%%%%%%%%%%%%%%%%%%%%%%%%%%%%%%%%%%%%%%%%%%%%%%%%%%%%%%%%
%%%%%%%%%%%%%%%%%%%%%%%%%%%%%%%%%%%%%%%%%%%%%%%%%%%%%%%%%%%%%%%%%%%%%%%%%%%%%%%%%%%%%%%%%%%

\begin{figure*}[ht!]
\begin{center}
\includegraphics[scale=1.1]{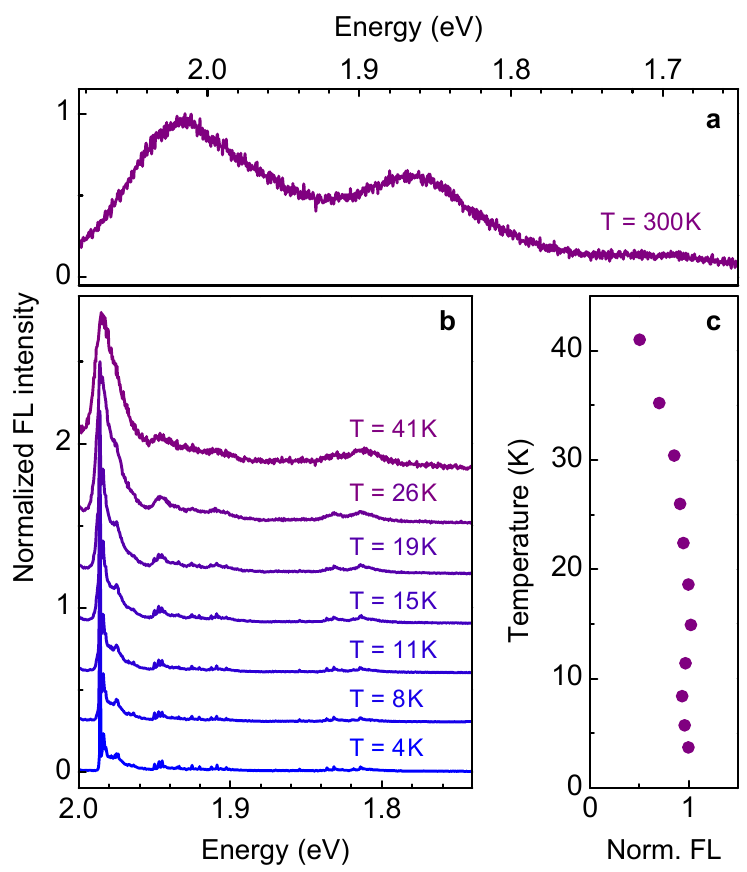}
\vspace{-15pt} \caption{\textbf{a}, Normalized fluorescence
spectra of a contaminant in anisole-based PMMA measured at $300$~K
(reproduced from Fig.~1f of the main text) and \textbf{b}, of a
similar fluorescence hotspot for selected temperatures ranging
from $4$~K to $41$~K (the traces were offset for clarity).
\textbf{c}, Total fluorescence normalized to the maximum value at
$4$~K as a function of temperature. All data were measured with
excitation at $532$~nm.} \label{si_qeprop}
\end{center}
\end{figure*}

The evolution of a typical type~1 hotspot spectrum with
temperature in \mbox{sample A} (anisole-based PMMA) is shown in
Fig.~\ref{si_qeprop}. The ZPL and the vibronic satellites
broadened upon heating from $4$ to $41$~K (Fig.~\ref{si_qeprop}b),
and the overall FL decreased gradually for temperatures above
$\sim 25$~K (Fig.~\ref{si_qeprop}c).  Both the initial intensity
and the spectral lineshape were recovered upon successive cooling
back to $4$~K. The trend of thermal broadening as in
Fig.~\ref{si_qeprop}b eventually results in significant spectral
overlap of the ZPL and vibronic satellites at room temperature
with predominant contributions from the vibronic group around
$\sim 155$~meV and its second harmonic around $\sim 310$~meV
red-shifts (Fig.~\ref{si_qeprop}a).

%%%%%%%%%%%%%%%%%%%%%%%% FIG %%%%%%%%%%%%%%%%%%%%%%%%%%%%%%%%%%%%%%%%%%%%%%%%%%%%%%%%%%%%%%
%%%%%%%%%%%%%%%%%%%%%%%%%%%%%%%%%%%%%%%%%%%%%%%%%%%%%%%%%%%%%%%%%%%%%%%%%%%%%%%%%%%%%%%%%%%

\subsection{Quantum yield estimate}\label{quantum_yield}
\vspace{-8pt}

The FL quantum yield of an emitter is given by the ratio of
emitted photons to absorbed photons per unit time. In our
experiments we estimate the quantum yield of single type~1
fluorescent hotspots in PMMA films formed with anisole-based
solutions by scaling the FL intensity to a photostable emitter
with known optical properties. We used single nitrogen-vacancy
(NV) color centers in bulk diamond that exhibit a quantum yield of
$\Phi_{\mathrm{NV}} \simeq 70\%$ \cite{Jelezko2006} and a dipole
averaged absorption cross-section of  $\sigma_{\mathrm{NV}} \simeq
3.1 \cdot 10^{-17}$~cm$^2$ \cite{Wee2007} for $532$~nm excitation.
With these quantities, the conversion cross-section of a
fluorescent hotspot, given by the product of the corresponding
quantum yield $\Phi_{\mathrm{FH}}$ and absorption cross-section
$\sigma_{\mathrm{FH}}$, is determined as:
\begin{equation*}
\label{eq_quantumyield} \Phi_{\mathrm{FH}} \, \sigma_{\mathrm{FH}}
= \frac{I_{\mathrm{FH}}}{I_{\mathrm{NV}}} \cdot
\frac{\tau_{\mathrm{FH}}}{\tau_{\mathrm{NV}}} \cdot
\frac{\Omega_{\mathrm{NV}}}{\Omega_{\mathrm{FH}}} \cdot
\Phi_{\mathrm{NV}} \, \sigma_{\mathrm{NV}}.
\end{equation*}
Here, the term $I_{\mathrm{FH}} / I_{\mathrm{NV}}$ is the FL
intensity of a hotspot scaled to the FL of a single NV center for
the same excitation power in the linear response regime of both
emitters. This ratio ranged from $1.9$ to $2.4$ for room
temperature measurements and peaked at $\sim 13.2$ for cryogenic
temperatures of $3.1$~K and $4.2$~K. The emitters were excited
with continuous wave excitation at $532$~nm and circular
polarization to ensure averaging over the possible orientations of
the transition dipole moments. The factor $\tau_{\mathrm{FH}} /
\tau_{\mathrm{NV}}$ accounts for the different FL lifetimes of the
fluorescent hotspots and the NV centers with $\tau_{\mathrm{FH}}
\simeq 3.6$~ns and $\tau_{\mathrm{NV}} \simeq 12.9$~ns determined
experimentally. Finally, we also account for the difference in the
effective collection solid angles for fluorescent dipoles embedded
in different dielectric environments (PMMA and diamond) with
respective refractive indices via $\Omega_{\mathrm{NV}} /
\Omega_{\mathrm{FH}}$ which was close to $ \simeq 0.35$ in our
experiments.

With these values the conversion cross-section of a typical type~1
fluorescent hotspot excited at $532$~nm was in the order of $\sim
5 \cdot 10^{-18}$~cm$^2$ at room temperature and increased to
$\sim 3 \cdot 10^{-17}$~cm$^2$ at cryogenic temperatures. Using
$\sigma_{\mathrm{FH}} \simeq 1 \cdot 10^{-16}$~cm$^2$ as a typical
absorption cross-section of the second absorption band of common
polycyclic hydrocarbon compounds \cite{Berlman1971}, we obtain an
estimate for the fluorescence quantum yield $\Phi_{\mathrm{FH}}$
of $\sim 5 \%$ at room temperature and up to $30 \%$ at cryogenic
temperatures. We obtained similar values, both at room and
cryogenic temperatures, from scaling the quantum yield of
fluorescent hotspots in anisole-based samples to fluorescence
characteristics of individual TDI molecules.

%\bibliographystyle{apsrev}
%\bibliography{bibliography_SI}

%%%%%%%%%%%%%%%%%%%%%%%%%%%%%%%%%%%%%%%%%%%%%

\end{document}